# MODELING OF STOCK RETURNS AND TRADING VOLUME

TAISEI KAIZOJI[1]

Graduate School of Arts and Sciences,

International Christian University,

3-10-2 Osawa, Mitaka, Tokyo 181-8585, Japan

## Abstract

In this study, we investigate the statistical properties of the returns and the trading volume. We show a typical example of power-law distributions of the return and of the trading volume. Next, we propose an interacting agent model of stock markets inspired from statistical mechanics [24] to explore the empirical findings. We show that as the interaction among the interacting traders strengthens both the returns and the trading volume present power-law behavior.

## 1. Introduction

Over half a century, a considerable number of researches have been made on trading volume and its relationship with asset returns [1-12]. Although the existence of the relationships between trading volume and future prices is inconsistent with the weak form of market efficiency [10], the analysis of the relationship has received increasing attention from researchers and investors. One of the causes of great attention to the relationship is that many have considered that price movements may be predicted by trading volume.

The researchers in a new field of science called 'econophysics' have worked on this problem from a slightly different angle. In the literature of econophysics [13-20], most studies on price fluctuations and trading volume have focused primarily on finding some universal characteristics which are often observed in complex systems with a large number of interacting units, such as power laws, and have modeled the statistical properties observed. Generally speaking, studies on price-volume relations tend to be very data-based, and the models are more statistical than economic in character.

[1] The corresponding author: Taisei Kaizoji, International Christian University, 3-10-2 Osawa, Mitaka, Tokyo 181-8585, Japan. E-mail: kaizoji@icu.ac.jp

The aims of this study are two-fold. We first investigate the statistical properties of returns and trading volume using a database that records daily transaction for securities listed in the Tokyo Stock Exchange from 1975 to 2002. As a typical example of company, we take *Fujita Corporation* who is a middle-size construction company. We find that the probability distributions of returns and of trading volume follow power-laws. To give an explanation on these findings, we next study a model that expresses trading volume and its relationship with asset returns. Our previous work [21-24] proposed interacting-agent models of price fluctuations in stock markets. In these studies we applied the so-called Ising models [25,26], which is a well-known model in statistical mechanics, to stock markets, and described the interaction of agents. In this paper we utilize our model [24], and formulate the relationship between returns and trading volume. In the model [24] the stock market is composed of the two typical groups of traders: the fundamentalists who believe that the stock price will be equal to the fundamental value [29], and the interacting traders who tend to get influenced by the investment attitude of other traders. We derive the market-clearing prices and the trading volume from demand for and supply of shares of a stock. We show that the probability distributions of returns and trading volume generated by computer simulations of the model have power-law tails when the interaction among the interacting traders, the so-called *conformity effect*, strengthens.

The rest of the paper is organized as follows. The next section reviews briefly the empirical findings. Section 3 describes the interacting agent model, and section 4 shows results of computer simulations, and section 5 gives concluding remarks.

## 2. The empirical results

In this section, we examine the statistical properties of the probability distributions of returns and of trading volume using a database that records daily transaction for all securities listed in the Tokyo Stock Exchange. We use the daily data for the closing price and the trading volume on the company in the 28-year period from January 1975 to January 2002 when corresponds to Japan's asset bubble and eventual burst. As a typical example of companies, we selected *Fujita Corporation* (Ticker: 1806), who is a middle-size construction company in Japan. We find that the probability distributions of returns and of trading volume on the company present power-law decay. Return is defined as the logarithmic price change of the stock from one day's close to the next day's close, and trading volume is defined as the number of shares traded in a trading day. Figure 1 (a) and Figure 1 (b) display the time series of returns and the probability distribution of returns drawn from the corresponding time series in Figure 1 (a). The time series of returns has clustered volatility that consists of laminar phases and chaotic bursts. Figure 1 (b) displays the semi-log plot of the probability distribution of the returns, and indicates that the tails of the return distribution have apparently longer than tails of an exponential distribution[2]. Figure 1(b) shows that the tails of return distribution are fit approximately by a power law, $P(|R|>x) \sim x^{-\alpha}$, where $R$ denotes returns, and $\alpha$ the power-law exponent.

---

[2] If the tails of the probability distribution can be approximated by a liner line, then

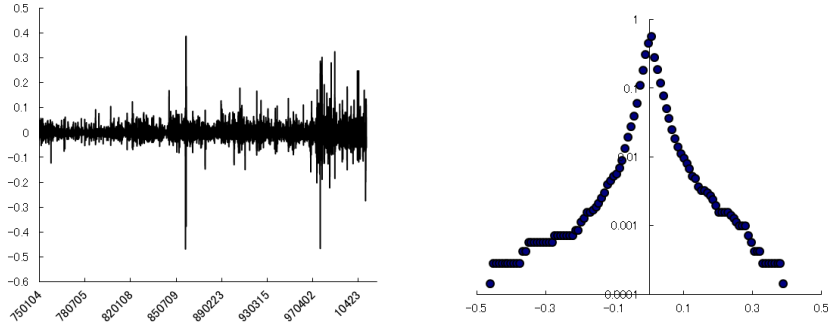

(a) The returns for Fujita      (a) The probability distribution of the returns for Fujita

Figure 1: The time series of (a) returns and (b) the probability distribution of returns for Fujita from January 1975 to January 2002. The tails of the probability distribution of returns displays power-law decay.

We next analyze the statistics of the trading volume. Figure 2 (a) and Figure 2 (b) show the time series of the trading volume, and the semi-log plot of the probability distribution of the trading volume calculated by the corresponding time series in Figure 2 (a). We also find that the probability distribution of the trading volume in Figure 2(b) also displays a power-law behavior.

To investigate the reason why power-laws for returns and trading volume are observed, we present an interacting agent model in the following section.

---

the probability distributions follow an exponential function.

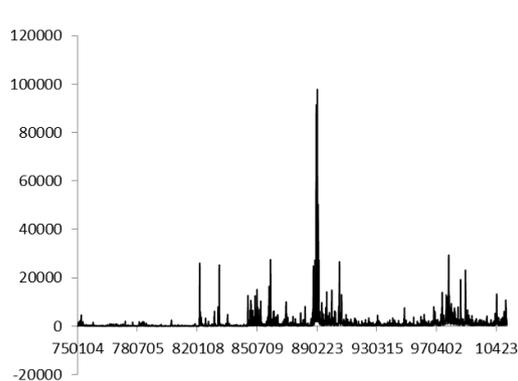
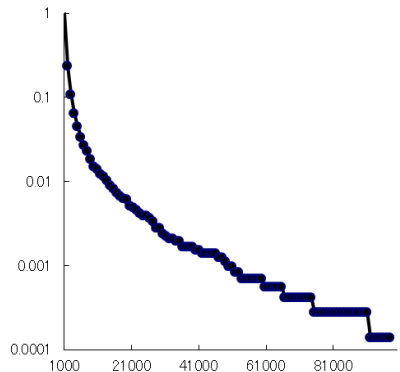

(a) The trading volume for Fujita

(b) The probability distribution of the returns for Fujita

Figure 2: The time series of (a) the trading volume and (b) the probability distribution of the trading volume for Fujita Corporation from January 1975 to January 2002. The tail of the probability distribution of the trading volume displays power-law decay.

## 3. An interacting-agent model

In this section we introduce an interacting-agent model proposed in [24]. We consider a stock market where shares of a stock are traded at price, $S_j$, per share. Two groups of traders with different trading strategies, *interacting traders* and, *fundamentalists* participate in the trading. The number of fundamentalists, $m$, and the number of interacting traders, $n$, are assumed to be constant. The model is designed to describe movements of the price of the stock, and of the trading volume over a trading day. In the following, a more precise account of the decision making of each trader type is given.

### 2.1. Interacting traders

An interacting trader is described in the literature of financial research as noise trader. Noise trader is frequently regarded as a trader who trade randomly [28].

However, an interacting trader is not idiot but makes decisions regarding buying or selling based on perceived market moods rather than fundamentals. An interacting trader buys when the others seem to be buying and sells when the others seem to be selling.

Each interacting trader is labeled by an integer $i, (i=1,2...,n)$. The investment attitude of interacting trader $i$ is represented by the random variable, $u_i$, and is defined as follows. If interacting trader $i$ is a buyer of shares during a given day, then $u_i = +1$, otherwise he/she sells shares, and then $u_i = -1$. Consider that the investment attitude $u_i$ is updated with transition probabilities. We adopt the following functions of the average investment attitude over interacting traders $X = \frac{1}{n}\sum_{i=1}^{n} u_i$ as the transition probabilities.

$$W_\uparrow(X) = \frac{1}{1+\exp(-2\phi X)} \quad \text{(transition from seller to buyer)} \quad (1)$$

$$W_\downarrow(X) = \frac{1}{1+\exp(2\phi X)} \quad \text{(transition from buyer to seller)} \quad (2)$$

where $\phi$ denotes intensity of interaction among interacting traders. We assume that $\phi$ vary randomly in time. If more than a half number of interacting traders are buyers, then the average investment attitude $X$ is positive, and if more than a half number of interacting traders are sellers, then the average investment attitude $X$ is negative. As the average investment attitude $X$ increases, the transition probability from a seller to a buyer $W_\uparrow(X)$ is higher, and at the same time, the transition probability from a buyer to a seller $W_\downarrow(X)$ is lower. Inversely, as the average investment attitude $X$ decreases, the transition probability from a buyer to a seller $W_\downarrow(X)$ is higher, and the transition probability from a seller to a buyer $W_\uparrow(X)$ is simultaneously lower.

One can derive[3] a following dynamic equation of the average investment attitude $X$ from the above transition probabilities (1)-(2):

$$dX(t) = K(X)dt + \sqrt{Q(X)}\,dW(t), \tag{3}$$

where $K(X) = \tanh(\phi X)$, $Q(X) = \dfrac{2}{n}[1 - \tanh(\phi X)]$, and $W(t)$ is the Wiener process. To perform the numerical simulation, we consider a discretized version of the stochastic differential equation (3)

$$X_{j+1} = X_j + K(X_j)\Delta t_j + \sqrt{Q(X_j)}\,\Delta W_j \tag{4}$$

Here, $X_j = X(t_j)$, $\Delta t_j = t_{j+1} - t_j$, and $\Delta W_j = W(t_{j+1}) - W(t_j)$. $\Delta W_j$ is the increment of the Winner process or a white noise. Following [24], we assume that intensity of the interaction changes randomly in time $j$, and furthermore, define the randomness as $\phi_j = \rho \Delta W_j$ where the parameter $\rho$ is a constant, and $\Delta W_j$ is the increment of the Winner process. We name the parameter, $\rho$, which describes the intensity of interaction among interacting traders, the *conformity effect*.

Each interacting trader chooses to either buy or sell shares, and is assumed to trade a fixed amount, $b$, of the shares in a day. Then the interacting-traders' *excess demand* for shares is defined as

$$Q_j^I = b n X_j. \tag{5}$$

## 2.2. Fundamentalists

In finance theories, the fundamental value, $S_j^*$, of the firm is estimated by forecasting future cash flows and discount rates [29]. Fundamentalists are assumed to have fundamental valuation techniques to find the fundamental value, $S_j^*$, and have information of the fundamentals of the firm. If the price, $S_j$, is below the fundamental value, $S_j^*$, a fundamentalist tends to buy shares because he estimates the stock to be undervalued in a stock market, and if the price, $S_j$, is above the fundamental value, $S_j^*$, a fundamentalist tends to sell shares. Hence we assume that fundamentalists' demand function is given by:

$$Q_j^f = a\,m\left(\ln S_j^* - \ln S_j\right). \tag{6}$$

where $m$ is the number of fundamentalists, and $a$ parametrizes the reaction on the discrepancy between the fundamental value and the market price.

---

[3] For the detail deviation of equations (3), and (4), see [24], and [27].

### 2.3. Market clearing price

The mechanism of trading in stock exchanges differs from one exchange to another. However, the price movements in any trading mechanism are governed by the forces of demand and supply. We here consider here an auction market which a *market maker* mediates the trading, and matches buyers with sellers at a market clearing price. The market transaction is performed when the buying orders are equal to the selling orders. The balance of demand and supply is written as

$$Q_j^f + Q_j^I = a\,m\left[\ln S_j^* - \ln S_j\right] + b\,n\,X_j = 0. \tag{7}$$

Hence the market price is calculated as

$$\ln S_j = \ln S_j^* + \lambda X_j \tag{8}$$

where $\lambda = \dfrac{bn}{am}$. Using the price equation (7), we can categorize the market situations as follows. If $X_j = 0$, the market price, $S_j$, is equal to the fundamental value, $S_j^*$. If $X_j > 0$, the market price, $S_j$, exceeds the fundamental value, $S_j^*$, (*bull* market regime). If $X_j < 0$, the market price, $S_j$, is less than the fundamental value, $S_j^*$, (*bear* market regime). The price changes are defined as

$$R_j = \ln S_j - \ln S_{j-1} = \left(\ln S_j^* - \ln S_{j-1}^*\right) + \lambda\left(X_j - X_{j-1}\right) \qquad (9)$$

Changes in the market price depend on changes in the interacting-traders' orders, and changes in fundamental value. We here assume that changes in the fundamental price $\left(\ln S_j^* - \ln S_{j-1}^*\right)$ are unpredictable and random[4]. This is a logical entailment of a well-known efficient market hypothesis [30]. In the efficient market, the price, $S_j$, is basically equal to the fundamental value, $S_j^*$. However, in this model, market clearing price, $S_j$, depends on not only fundamental value, $S_j^*$, but also interacting traders' orders. Hence, the efficient market does not hold. From the market clearing condition (7), the trading volume, which is defined as the number of shares traded in a trading day, is calculated by the following,

$$V_j = bn\left(\dfrac{1 + \left(|X_j|\right)}{2}\right). \qquad (10)$$

## 3. Simulations

---

[4] We assume that the term $\left(\ln S_j^* - \ln S_{j-1}^*\right)$ is a white noise.

In this section we perform computer simulations of the model (9)-(10) proposed in the preceding section. Hereafter, the model will be simulated numerically with the parameters given by $\Delta t_j = 0.1$, $\lambda = 1$, and the fixed number of interacting traders, $n = 100,000$. We investigate statistical properties of the dynamics of returns, $R_j$, and of trading volume, $V_j$, for three different values of the conformity effect, which indicates intensity of interaction among the interacting traders, $\rho = 0.1$, 2 and 8. Figure 3 (a) and Figure 3 (b) display the typical evolutions of the returns, $R_j$, and of the trading volume, $V_j$, for a weak conformity effect, $\rho = 0.1$. Both the time series have only narrow Gaussian fluctuations. In Figure 3 (c) and in Figure 3 (d) we show the probability distributions of returns, $R_j$, and of trading volume, $V_j$, for $\rho = 0.1$. The tails of both the probability distributions of returns, $R_j$, and of trading volume, $V_j$, can be approximated by normal distributions.

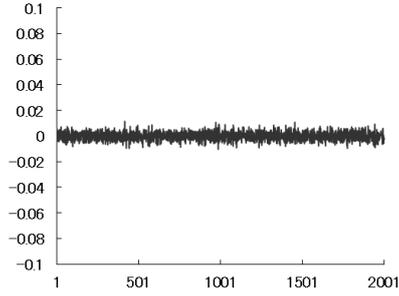
(a) Time series of the returns for $\rho = 0.1$

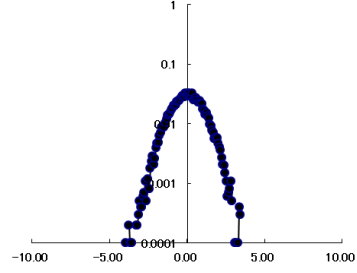
(c) The probability distribution of the returns for $\rho = 0.1$

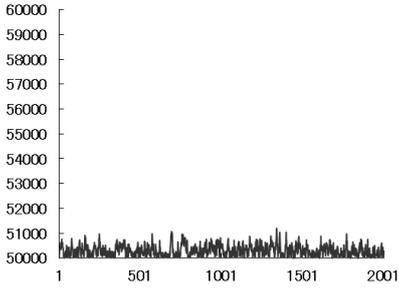
(b) Time series of the trading volume for $\rho = 0.1$

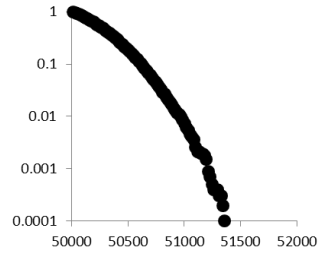
(d) The probability distribution of the trading volume for $\rho = 0.1$

Figure 3: A weak conformity effect $\rho = 0.1$. The time series of (a) returns and of (b) trading volume are generated from the model with $\rho = 0.1$, $\Delta t_j = 0.1$, $\lambda = 1$, and $n = 100,000$. The semi-log plots of probability distributions of (c) returns and of (d) trading volume for $\rho = 0.1$. When we assume a weak conformity effect, $\rho = 0.1$, the tails of the probability distributions of returns and of trading volume follow normal distributions.

Figure 4 (a) and Figure 4 (b) display the time series of returns, $R_j$, and trading volume, $V_j$, for a moderate conformity effect, $\rho = 2$. Both the time series are more volatile than the time series of returns, $R_j$, and of trading volume, $V_j$, for the very weak conformity effect $\rho = 0.1$. The tails of the probability distributions of returns, $R_j$, and of trading volume, $V_j$, for $\rho = 2$ can be approximated by exponential distributions (Figure 4 (c) and in Figure 4 (d)).

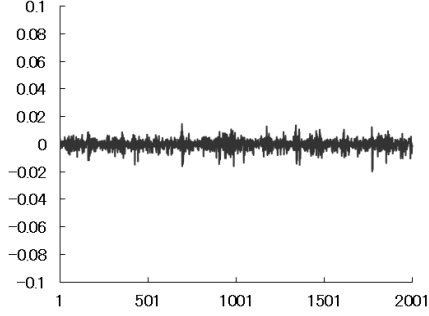

(a) Time series of the returns for $\rho = 2$

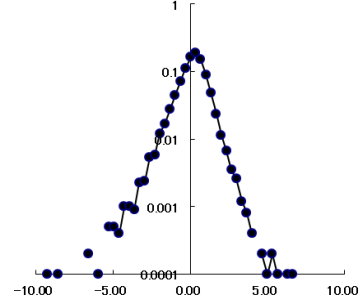

(c) The probability distribution of the returns for $\rho = 2$

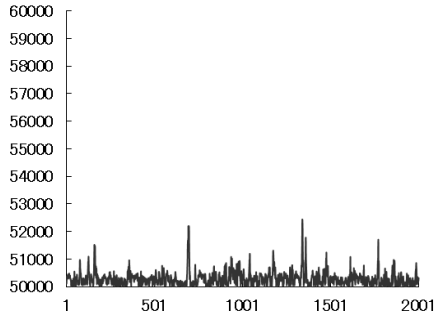

(b) Time series of the trading volume for $\rho = 2$

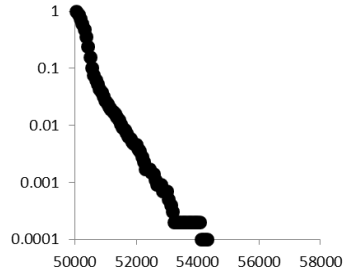

(d) The probability distribution of the trading volume for $\rho = 2$

Figure 4: A moderate conformity effect $\rho = 2$. The time series of (a) returns and of (b) trading volume are generated from the model with $\rho = 2$, $\Delta t_j = 0.1$, $\lambda = 1$, and $n = 100,000$. The semi-log plots of probability distributions of (c) returns and of (d) trading volume for $\rho = 2$. When we assume a moderate conformity effect, $\rho = 2$, the tails of the probability distributions of returns and of trading volume follow exponential distributions.

Let us raise the value of the parameter $\rho = 8$ from $\rho = 2$. Figure 5 (a) and Figure 5 (b) display the time series of returns, $R_j$, and trading volume, $V_j$, for a strong conformity effect, $\rho = 8$. Both the time series of the returns, $R_j$, and the trading volume, $V_j$, for $\rho = 8$ have clearly clustered volatility, and consist of laminar phases and chaotic bursts. For a strong conformity effect, $\rho = 8$ the tails of the probability distributions of returns, $R_j$, and trading volume, $V_j$, shows power-law decay (Figure 5 (c) and Figure 5 (d)).

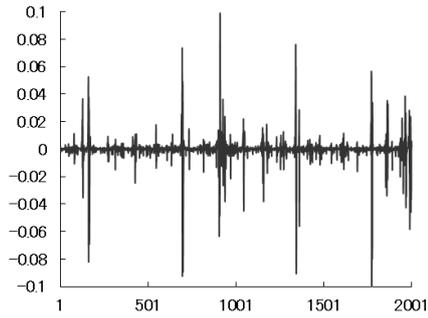

(a) Time series of the returns for $\rho = 8$

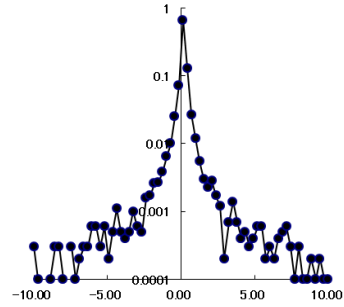

(c) The probability distribution of the returns for $\rho = 8$

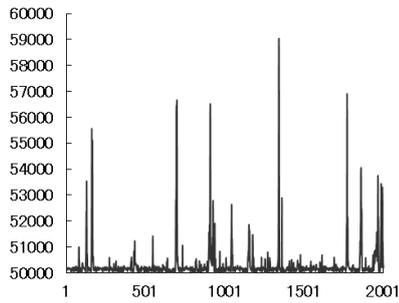

(b) Time series of the trading volume for $\rho = 8$

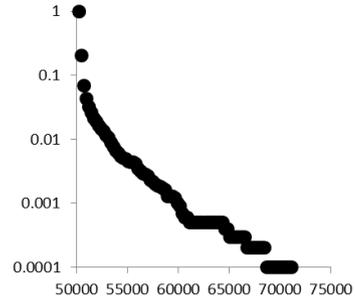

(d) The probability distribution of the trading volume for $\rho = 8$

Figure 5: A strong conformity effect, $\rho = 8$. The time series of (a) returns and of (b) trading volume are generated from the model with $\rho = 8$, $\Delta t_j = 0.1$, $\lambda = 1$, and $n = 100,000$. The semi-log plots of probability distributions of (c) returns and of (d) trading volume for $\rho = 8$. When we assume a strong conformity effect, $\rho = 8$, the tails of the probability distributions of returns and of trading volume follow power-law distributions.

It can be seen that, as the conformity effect $\rho$ increases, clustered volatility comes to be observed more definitely. With strengthening the conformity effect, $\rho$, the shape of the probability distributions of returns, $R_j$, and of trading volume, $V_j$, change from Gaussian distributions to power-law distributions. To sum up, as the interaction among the interacting traders strengthens, the returns and the trading volume present a power-law behavior.

## 4. Concluding remarks

In this paper we present an interacting agent model of a stock market which is modeled from the point of view of statistical mechanics. In our model the stock market is composed of the two groups of traders: the fundamentalists and the interacting traders. The point of the results shown by the simulation is that intensity of the interaction among the interacting traders is a key factor which generates power-law distributions of returns and of trading volume.

The question, which has been touched in introduction but not explored, is analysis of the relationships between returns and trading volume. It needs further consideration.

## 5. Acknowledgement

This research was supported by the Grant-in-Aid, No. 25380404 from the Japan Society for the Promotion of Science (JSPS).